\documentclass[10pt,
aps,physrev,
twocolumn,
showkeys,
superscriptaddress
]{revtex4-2}

\usepackage{amsmath}
\usepackage{graphicx}
\usepackage{xcolor}
\usepackage{placeins}
\usepackage{float}
\usepackage{hyperref}

\begin{document}


\title{Control of eigenmode localization and coupling anisotropy by multiscan femtosecond laser writing}


\author{S.A. Zhuravitskii}
\email{zhuravicky.sa15@physics.msu.ru}
\affiliation{
 Quantum Technology Centre, Faculty of Physics, M.V. Lomonosov Moscow State University, 1 Leninskie Gory Street, Moscow 119991, Russia}

\author{N.S. Kostyuchenko}
\affiliation{
 Quantum Technology Centre, Faculty of Physics, M.V. Lomonosov Moscow State University, 1 Leninskie Gory Street, Moscow 119991, Russia}

\author{N.N. Skryabin}
\affiliation{
 Quantum Technology Centre, Faculty of Physics, M.V. Lomonosov Moscow State University, 1 Leninskie Gory Street, Moscow 119991, Russia}
\affiliation{
 Russian Quantum Center, 30 Bolshoy bul'var building 1, Moscow 121205, Russia}

\author{I.V. Dyakonov}
\affiliation{
 Quantum Technology Centre, Faculty of Physics, M.V. Lomonosov Moscow State University, 1 Leninskie Gory Street, Moscow 119991, Russia}
\affiliation{
 Russian Quantum Center, 30 Bolshoy bul'var building 1, Moscow 121205, Russia}

\author{A.A. Korneev}
\affiliation{
 Quantum Technology Centre, Faculty of Physics, M.V. Lomonosov Moscow State University, 1 Leninskie Gory Street, Moscow 119991, Russia}

\author{A.A. Kalinkin}
\affiliation{
 Quantum Technology Centre, Faculty of Physics, M.V. Lomonosov Moscow State University, 1 Leninskie Gory Street, Moscow 119991, Russia}

\author{S.S. Straupe}
\affiliation{
 Quantum Technology Centre, Faculty of Physics, M.V. Lomonosov Moscow State University, 1 Leninskie Gory Street, Moscow 119991, Russia}
\affiliation{
 Russian Quantum Center, 30 Bolshoy bul'var building 1, Moscow 121205, Russia}
 
\author{S.P. Kulik}
\affiliation{
 Quantum Technology Centre, Faculty of Physics, M.V. Lomonosov Moscow State University, 1 Leninskie Gory Street, Moscow 119991, Russia}

\date{\today}

\begin{abstract}
The multiscan method is widely used in femtosecond laser writing of waveguide systems to increase the refractive index contrast and to control the eigenmode shape. Here, we investigate multiscan writing with a constant effective writing speed and show that the displacement between adjacent scans provides an additional degree of freedom for controlling eigenmode confinement. By optimizing the multiscan parameters, the effective mode area can be reduced, enabling an up to eightfold increase in fabrication speed compared to conventional single-scan writing. We further demonstrate that careful tuning of the multiscan geometry substantially reduces the anisotropy of inter-waveguide coupling. These findings are applied to the fabrication of three-dimensional waveguide arrays operating at wavelengths of $808\,\textrm{nm}$ and $1450\,\textrm{nm}$, where reduced coupling anisotropy is achieved. The proposed approach may be useful for three-dimensional waveguide arrays employed in topological photonics experiments, where low coupling anisotropy is desired.
\end{abstract}


\maketitle


\section{\label{sec:Introduction} Introduction}


    The femtosecond laser direct writing (FLDW) of waveguides in a volume of transparent materials, first demonstrated in 1996 \cite{davis1996writing}, is widely used for the fabrication of integrated optical waveguide structures. During the FLDW process, material modification occurs in the focal region, which makes it possible to manufacture complex three-dimensional integrated optical circuits. This feature is extensively employed in optical communications \cite{thomson2011ultrafast, jia2014monolithic,  maclachlan2015development, cai2022femtosecond} and quantum optical experiments \cite{hoch2022reconfigurable, feng2023direct, zhou2024multi}. 3D capabilities of FLDW take a special place in topological photonics, as they enable the creation of waveguide arrays with complex geometries \cite{rechtsman2013photonic, maczewsky2020nonlinearity, li2023observation, arkhipova2023observation, yan2024photonic}, which are capable of mimicking solid-state systems with topological properties.

    It is often required that the waveguides in a 3D structure be identical. However, this requirement is usually not met in the FLDW process, because waveguide parameters depend on the writing depth due to spherical aberrations occurring when the beam is focused through the glass-air interface. To reduce spherical aberrations, several methods may be employed, such as writing with an immersion objective \cite{wang2024precise}, correction of aberrations using spatial light modulator (SLM) \cite{salter2014exploring, tang2018experimental}, or writing with a low numerical aperture \cite{maczewsky2020nonlinearity, li2023observation}.

    In the case of topological waveguide arrays, the symmetry of the system can play an important role. Since the cross-section of laser-written waveguides is usually elliptical, the coupling between the waveguides is anisotropic (depends on their relative orientation) \cite{szameit2007coupling}. Sometimes, the anisotropy can be neglected by properly aligning the symmetry axes of the entire array and the individual waveguide \cite{arkhipova2023observation}. However, this strategy cannot eliminate the anisotropy in lattices with multiple nonequivalent coupling directions imposed by system symmetry \cite{ren2023observation, zhong2024observation}. This anisotropy introduces some difficulties in experiments with topological systems. First, the anisotropy of couplings leads to narrowing of the topological gap and, as a consequence, reduces localization of topological states. Second, in systems with rotational symmetry, vortex states \cite{huang2024vortex} can exist, which are a superposition of degenerate eigenmodes. However, the anisotropy of the couplings breaks the degeneracy and makes it impossible to observe vortex states \cite{kireev2026observation}. There are several approaches to reduce the ellipticity of the waveguide cross-section (and therefore the anisotropy of the inter-waveguide coupling): writing waveguides in a cumulative heating regime \cite{arriola2013low, hoch2022reconfigurable, noh2017experimental}, shaping the writing beam using a mechanical slit \cite{ams2005slit, tan2025slit}, cylindrical optics \cite{osellame2003femtosecond} or SLM \cite{salter2012adaptive, li2021circular}, as well as manipulating the cross-section of a waveguide using a multiscan technique \cite{nasu2005low, liu2004waveguide}.

    The latter approach is a simple and effective method to change waveguide properties. Multiple overlapping scans increase the contrast of the refractive index, \cite{tan2020fabricating}, while additional displacement between consecutive scans enables control of the eigenmode profile \cite{nasu2005low, dyakonov2016low, wang2024precise, chen20243d}. The use of multiscan is combined with spherical aberration control \cite{sun2024fast, ross2024low, skryabin2024femtosecond} and management of mechanical stress inside the material \cite{lee2021low, sun2020control}. It is widely used to reduce coupling and bending loss \cite{tan2020fabricating, tan2022effectively, ehrhardt2023high, skryabin2024femtosecond, xiao2025low}, manipulate spatial modes and their directional coupling \cite{wang2024precise}, dynamically transform spatial \cite{sun2024fast, heilmann2018tapering, li2023photon} and polarization modes \cite{sun2022chip}, and control polarization anisotropy of waveguides \cite{li2021circular, wang2023ultralow, ehrhardt2021exploring, memeo2026tailoring}. However, in previous studies there has been insufficient attention paid to the impact of the distribution of laser pulses across the waveguide cross-section on the eigenmode shape. Also, to the best of our knowledge, the use of multiscan technique  to reduce the anisotropy of inter-waveguide coupling has not been studied before.
    
    In our work, we first fixed the exposure conditions for each waveguide, systematically investigated the effect of the number of scans and their relative displacement on eigenmode shape, and compared the results with a single-scan waveguide. This allowed us to accelerate the writing process by a factor of eight and adapt the initial writing regime to a larger wavelength range. We also demonstrated that multiscan can effectively reduce the anisotropy of inter-waveguide coupling. Finally, we illustrated the potential of the multiscan approach for fabricating uniform 3D waveguide arrays.

\section{\label{sec:Methods} Methods}

\subsection{FLDW parameters}
    The main parameters of the FLDW setup are listed below and were preserved in all experiments. We used the frequency doubled 
    ytterbium-doped amplified femtosecond fiber laser (Avesta Antaus) with a wavelength of $515\,\textrm{nm}$, repetition rate of $1\,\textrm{MHz}$ and pulse duration of $215\,\textrm{fs}$. The circularly polarized laser beam was focused by an aspheric lens with $\text{NA} = 0.4$. An aperture of a lens was partially filled giving an effective $\text{NA}_{eff} \approx 0.2$. The pulse energy, writing speed, number of scans, and its relative displacement are specified in each experimental section. A $5~\textrm{cm}$ long fused silica glass sample (JGS1) was placed on a precision positioning system (AeroTech, FiberGlide 3D), which provided a maximum translation speed of $64\,\textrm{mm/s}$. The end faces of the sample were polished after the writing process. 

    We used a pulse energy of $270\,\textrm{nJ}$ and writing speed of $1\,\textrm{mm/s}$ for writing a reference single-scan waveguide (RWG). This writing regime corresponds to type-I modification (uniform refractive index change \cite{itoh2006ultrafast, poumellec2011modification, mishchik2013ultrafast}). The resulting waveguide exhibits mode-field diameters (MFDs) of approximately $15\,\mu\textrm{m}\times21\,\mu\textrm{m}$ (see Appendix~\ref{appendix:RWG}) and propagation loss of about $0.1\,\textrm{dB/cm}$ at $808\,\textrm{nm}$. Relatively low numerical aperture of the focusing optics also enables fabrication of nearly identical waveguides over a depth range of more than $400\,\mu\textrm{m}$ (see Supplemental materials of Ref.~\cite{ren2023observation} for details). Although the eigenmode of such waveguides is not well-suited for coupling with SM fiber, this writing regime has been found to be highly effective for applications involving 3D waveguide arrays and free-space coupling of light to optical chips \cite{ren2023observation, arkhipova2023observation}. The main disadvantage of such waveguides for the tasks considered is their intrinsic ellipticity and related inter-waveguide coupling anisotropy.

\subsection{Eigenmode characterization}
    The eigenmodes of the waveguides were characterized using a $808\,\textrm{nm}$ pigtailed diode laser (Thorlabs). The laser beam was collimated and coupled into/out of the waveguides using aspheric lenses. The focal length of the coupling lens was chosen to optimize the coupling efficiency through the RWG. The sample was placed on a six-axis positioner (Luminos). The output intensity distribution was imaged onto a CMOS beam profiler (Gentec). MFDs were measured at the $1/e^{2}$ level. Effective mode area (EMA) was estimated as $\textrm{EMA} = \pi \frac{\textrm{MFD}_1 \textrm{MFD}_2}{4}$, and the aspect ratio of the eigenmode is given by the ratio of a smaller MFD to a larger one. All measurements were conducted for a horizontally polarized mode.

    In order to verify whether the waveguide is single-mode, we coupled a laser into the waveguide and observed the output intensity distribution on a camera. Then, the sample was shifted orthogonally to the optical axis. The waveguide was classified to be single-mode if the intensity distribution on the camera remained unchanged and to be few-mode  otherwise.

    In the case of few-mode waveguides, MFDs were measured for the fundamental modes. To this end, the sample was aligned until the camera image displayed a single intensity maximum and the minimum achievable spot size — a condition that also corresponded to the highest transmission through the system. To ensure that such measurements yield correct results, the same procedure was repeated using an alternative focusing lens that produced a beam waist smaller than the waveguide cross-section. The results of the two measurements typically differed by no more than $5\%$.

\subsection{Loss measurement}

    To measure the losses in the waveguides, we used the same setup as that employed for eigenmode characterization. Insertion loss was calculated as $IL = -10 \lg \left(\frac{P_{out}}{P_{in}}\right)$, where $P_{in}$ and $P_{out}$ are the optical power launched into and transmitted through the waveguide, respectively. Alternatively, $IL = PL \cdot l_{wg} + CL + BL \cdot l_{bend} + 2\cdot FL$, where $PL$ is the propagation loss, $CL$ is the coupling loss, $BL$ is the bending loss, $FL$ is the Fresnel loss associated with reflection at the glass-air interface, $l_{wg}$ is the total waveguide length, and $l_{bend}$ is the length of the curved section. Fresnel loss ($FL$) is taken into account twice, since reflections occur at both the input and output end faces of the sample. Coupling loss ($CL$) was evaluated using the overlap integral between the electric field of the input beam waist $E_{in}$ and the waveguide mode $E_{wg}$:
    \begin{equation}
        \eta = \frac{|\iint E_{in} E_{wg}dx dy|^2}{\iint |E_{in}|^2 dx dy \; \iint |E_{wg}|^2 dxdy},
    \end{equation}
    then $CL = -10 \lg{\eta}$. $E_{in}$ and $E_{wg}$ were estimated as the square root of the corresponding intensity distribution captured by a beam profiler, which is valid both for the Gaussian beam waist and the fundamental waveguide mode. Fresnel loss was estimated as $FL = -10 \lg(1-R) \approx 0.16\,\textrm{dB}$, where  $R = \left(\frac{n_{glass} - n_{air}}{n_{glass} + n_{air}} \right)^2$ is the Fresnel reflection coefficient for normal incidence and $n_{glass} = 1.46$, $n_{air} = 1$ are the refractive indices of glass and air. When CL and FL are known, PL may be estimated for straight waveguides using the equation $PL = \frac{IL - CL - 2\cdot FL}{l_{wg}}$. Finally, BL in the curved waveguides can be estimated as $BL = \frac{IL - PL\cdot l_{wg} - CL - 2\cdot FL}{l_{bend}}$.
    
\subsection{Coupling constant estimation}
    
    The coupling constant between waveguides was numerically estimated based on a comparison of the discrete diffraction pattern (DDP) in the one-dimensional waveguide lattices. The evolution of laser radiation in an array of identical, equally spaced single-mode waveguides can be approximately described by a system of coupled-mode equations \cite{snyder2012optical}:
    \begin{equation}
        i \frac{\partial  a_{n}(z)}{\partial z} + 
        C \sum_{k \in S_{n}} a_{k}(z) = 0, \qquad n = \overline{1, N},
    \end{equation}
    where $a_n$ is the field amplitude in the waveguide with the corresponding index, $C$ is the coupling constant between nearest neighbors, $S_{n}$ is the set of the nearest neighbors to the $n^{th}$ waveguide, and the summation is carried out over all waveguides.

    To estimate the coupling constant, we considered the excitation of the central waveguide in a system and compared, using root mean square error (MSE), an experimentally obtained output intensity distribution $I_{n}^{exp}$ with a theoretically calculated one $I_{n}^{th} = |a_{n}|^{2}$ for different values of the coupling constant:
    \begin{equation}
        MSE = \frac{1}{N}\sum_{n=1}^N (I_n^{exp} - I_n^{th})^2.
    \end{equation}
    Experimental distributions $I_{n}^{exp}$ were obtained from the DDPs captured by a camera and further numerical integration over small areas around each waveguide.

    The coupling constant between two waveguides is commonly expressed through the overlap integral of the mode fields $e_1$ and $e_2$ (normalized to unit power) with the permittivity perturbation $\Delta\varepsilon_2$ induced by the neighboring waveguide \cite{snyder2012optical}:
    \begin{equation}
    C = \frac{\omega\varepsilon_0}{4} \int_{-\infty}^{\infty}\int_{-\infty}^{\infty} 
        \Delta\varepsilon_2(x,y) {e}_1(x,y) {e}_2(x,y)\,dx\,dy,
    \end{equation}
    where $\omega$ is the angular frequency of light and $\varepsilon_0$ is the vacuum permittivity. The laser-written waveguides usually have an elongated cross-section and an elliptical eigenmode, resulting in an angular dependence of the coupling constant $C$ on the tilt angle $\varphi$ between the elliptical cross-sections and the line connecting the waveguide centers. Following \cite{szameit2007coupling}, the experimental dependencies of the coupling constant $C$ on $\varphi$ were fitted to the polar form of an ellipse centered at the origin:
    \begin{equation}
        C(\varphi) = \frac{C_{min} \cdot C_{max}} { \sqrt{C_{max}^2 \cdot \cos^2(\varphi - \varphi_0) + 
        C_{min}^2 \cdot \sin^2(\varphi - \varphi_0)} },
    \end{equation}
    where $C_{max}$ and $C_{min}$ are the major and minor semiaxes, respectively. The coupling anisotropy was defined as $\Delta_{C} = (C_{max} - C_{min}) / C_{max}$.

\section{\label{sec:Experiment} Experiment and results}

\subsection{\label{sec:multiscan} Multiscan under fixed exposure}

    \begin{figure*}
        \includegraphics[width=1.0\linewidth]{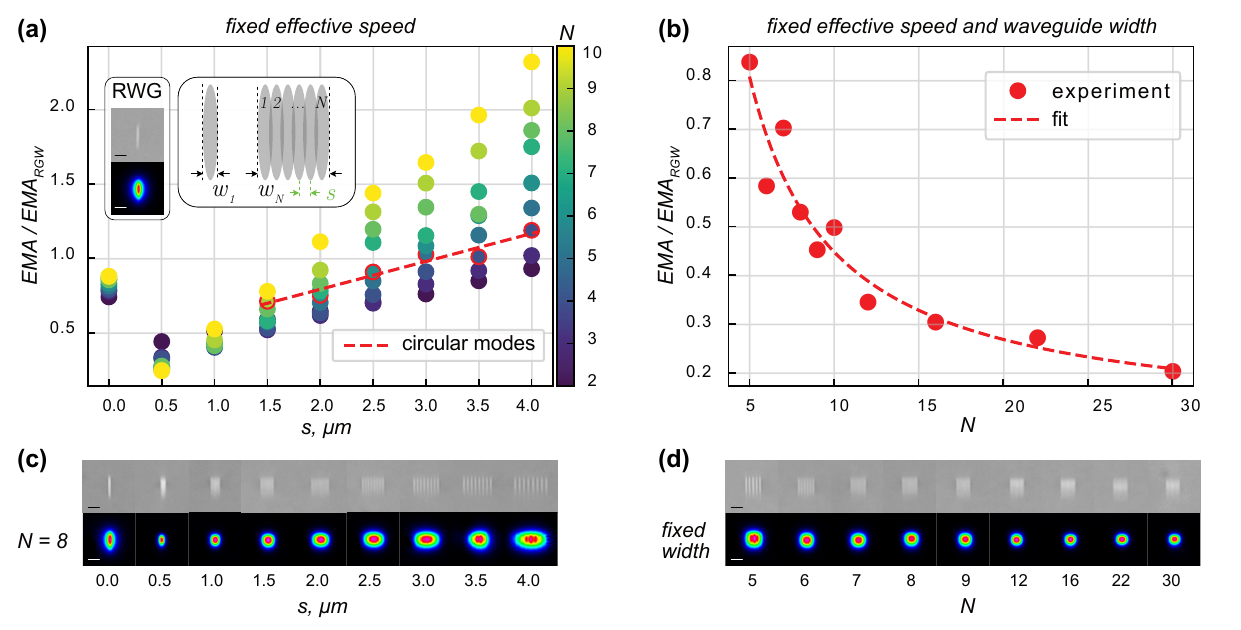}
        \caption{ \label{fig_1}
        (a) Dependence of the EMA, normalized to that of the RWG, on the displacement between scans. The color indicates the number of scans. The red dashed line indicates that the selected points (for waveguides with a nearly circular eigenmode) lie on a straight line. The insets show a bright-field (BF) microscope image of the RWG cross-section and its eigenmode (left), and the layout of the multiscan waveguide cross-section (right).
        (b) Dependence of the normalized EMA on the number of scans for waveguides with a fixed width $w_N = 12\,\mu\textrm{m}$. The dotted line shows the inverse-proportional fit.
        (c) BF images of the waveguide cross-section with varying displacement for the number of scans $N = 8$ and corresponding eigenmodes. The images are aligned with the $s$‑axis in panel (a).
        (d) BF images of the waveguide cross-sections shown in panel (b), together with the corresponding eigenmodes.
        Scale bars correspond to $10\,\mu\textrm{m}$.        
        }
    \end{figure*}

    In this section, we investigated the effect of the number of scans $N$ and their relative displacement $s$ on the eigenmode aspect ratio. We started with an RWG (see \hyperref[sec:Methods]{\textbf{Methods}}), which supports single mode with an aspect ratio $AR \approx 0.7$ and hence possesses anisotropic inter-waveguide coupling. Multiscan with a small displacement between individual scans allows changing the waveguide cross-section and, consequently, the eigenmode shape \cite{ehrhardt2023high, skryabin2024femtosecond}. Our goal was therefore to develop a writing regime of waveguides with reduced coupling anisotropy based on a multiscan approach. The multiscan approach without scan displacement leads to an increase in the induced refractive index $\Delta n$ compared to the original modification \cite{tan2020fabricating}. To prevent transition to the multimode regime during the multiscan process, we fixed the exposure conditions (namely, pulse energy and effective writing speed) to those of the RWG. In particular, the pulse energy was set to $E_p = 270\,\textrm{nJ}$, and the effective writing speed was $v_{eff} = v_{scan} / N = 1\,\textrm{mm/s}$, where $v_{scan}$ is the writing speed of an individual scan. Following this approach implies that the number of pulses per unit length of the waveguide remains constant, but they are redistributed over the cross-section and length of the waveguide.
    It is also worth noting that all waveguides in this section were written using the unidirectional scanning strategy (see Appendix~\ref{appendix:scan_ordering}).

    We considered waveguides with the number of scans $N$ from 2 to 10. For each $N$, the displacement between adjacent scans $s$ was chosen in the range from $0$ to $4\,\mu\textrm{m}$ and performed in the direction of the minor axis of a single modification (track), as shown in Fig.~\ref{fig_1}(a). This approach not only allows changing the aspect ratio of the eigenmode but also significantly affects its size, despite fixed exposure conditions for all waveguides.
    
    Fig.~\ref{fig_1}(a,c) shows the dependencies of EMA on relative displacement for each $N$, which can be explained based on the physical intuition provided below. For zero displacement, corresponding to complete overlap of adjacent scans, the amplitude of the induced refractive index $\Delta n$ is maximal, while the width of the modified area is minimal and approaches that of a single-scan modification. As displacement increases, the waveguide cross-section broadens, resulting in a higher V-number even though $\Delta n$ gradually diminishes (see Appendix~\ref{appendix:V_number}). Consequently, the EMA is reduced as long as there is significant overlap between adjacent tracks. As the displacement increases further and the overlap becomes negligible, $\Delta n$ remains constant, while the waveguide width continues to increase linearly with $s$, leading to corresponding EMA growth. For displacement values around $3\,\mu\textrm{m}$, the eigenmode becomes distorted due to the inhomogeneities inside the waveguide.
    
    When comparing the dependencies of EMA on displacement for different number of scans, three distinct regions can be identified on the graph. First, in the case of multiscan without displacement ($s = 0\,\mu\textrm{m}$), EMA is decreased in comparison to RWG. This region is discussed in more detail in Appendix~\ref{appendix:zero_displacement}. Second, the minimum EMA for all values of $N$ is observed near $s = 0.5\,\mu\textrm{m}$, and the lowest EMA value is achieved with the highest $N$. The optimum displacement at which this minimum EMA is attained depends on the number of scans (see Appendix~\ref{appendix:small_s_range}). Finally, for displacement values $s > 0.5\,\mu\textrm{m}$, the EMA grows as $N$ increases.
    
    Most waveguides with $s \neq 0\,\mu\textrm{m}$ were found to be few-mode, despite identical exposure conditions. This behavior originates from the large size of a single modification, which results in a pronounced rise of the V-number during multiscan. In these cases, the measurements are provided for the fundamental mode of the waveguide (\hyperref[sec:Methods]{\textbf{Methods}}). While many applications require single-mode waveguides, the transition to the multimode regime is not detrimental. First, single-mode behavior can be preserved by increasing the effective writing speed - a significant advantage when writing arrays of multiple waveguides. Second, increasing the operating wavelength reduces the V-number, thereby enabling single-mode operation at longer wavelengths. Both of these aspects will be addressed in the following sections.
    
    Broadening of the waveguide during multiscan leads to a higher aspect ratio of its eigenmode. For different number of scans $N$, the waveguides with nearly circular eigenmode were found to have similar widths $w_{N} = s_{N}\cdot(N-1) + w_{1} = w^{circ} \approx 12\,\mu\textrm{m}$, where $s_{N}$ is the corresponding scan displacement and $w_{1} \approx 1\,\mu\textrm{m}$ is the width of a single track. Because the number of scans $N$ was insufficient near $s \sim 0.5\,\mu\text{m}$ — the displacement at which the EMA reaches its minimum for all $N$ — no waveguides with nearly circular eigenmodes were observed in this range. We therefore wrote an additional series of fixed-width waveguides with $s_N = (w^{circ} - w_1)/(N-1)$ for $N$ values up to $30$. The dependence of EMA on $N$ is shown in Fig.~\ref{fig_1}(b,d). We observed an inversely proportional dependence of EMA on $N$ (or linear on $s$, since $s \cdot N = \textrm{const}$) across the entire range. This can be understood as follows. Increasing $N$ results in a more homogeneous pulse distribution across the waveguide cross‑section, thereby reducing the exposure per unit volume. Consequently, the refractive index volume (Appendix~\ref{appendix:V_number}) increases because the modification process in each unit volume remains far from saturation and thus contributes more efficiently to the cumulative change.

\subsection{\label{sec:speed-up} Acceleration of the writing process}

    \begin{figure}[b]
        \includegraphics[width=1.0\linewidth]{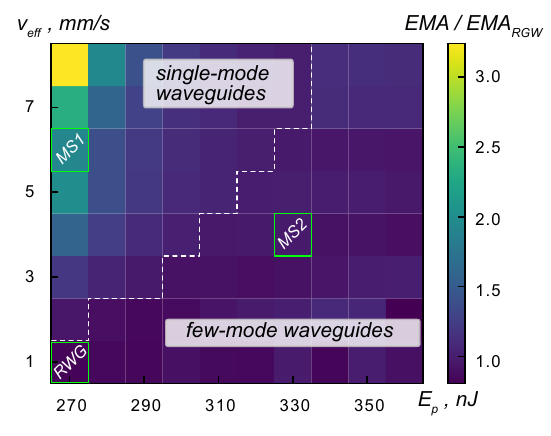}
        \caption{\label{fig_2}
        Heat map of the EMA, normalized to that of the RWG, as a function of the writing pulse energy and the effective writing speed for multiscan waveguides with $N = 8$ and $s = 1.5\,\mu\textrm{m}$. The white dashed line indicates the approximate boundary between the single-mode and few-mode regimes. For more information on the characterization of the guiding regime, see \hyperref[sec:Methods]{\textbf{Methods}}. 
        The green boxes mark the locations of the RWG, MS1, and MS2 waveguides, which are introduced in the following sections, in the $v_{eff}$-$E_{p}$ parameter space.
        } 
    \end{figure}
    
    In the previous section, we observed that multiscan writing with a fixed effective speed made the waveguide multimode for different numbers of scans and displacements. Here, we investigate the increase in writing speed to preserve single-mode operation. We fixed $N = 8$ and $s = 1.5\,\mu\textrm{m}$, corresponding to the waveguide with an eigenmode aspect ratio of $0.96$. Since the writing speed of a single scan $v_{scan} = v_{eff} \cdot N$ is limited by the maximum translation speed of the positioning system ($64\,\textrm{mm/s}$ for our setup), this allowed us to use the effective speeds up to $v_{eff} = 8\,\textrm{mm/s}$.

    Increasing the writing speed reduces $\Delta n$, thereby shifting the waveguide into the single-mode regime. Conversely, a higher pulse energy increases $\Delta n$, permitting higher effective writing speeds while maintaining single-mode operation. Furthermore, here and throughout the rest of this work, we used the bidirectional scanning strategy to eliminate the non-writing sample translations inherent to a unidirectional one (see Appendix~\ref{appendix:scan_ordering}).
    
    Fig.~\ref{fig_2} shows the heat map of EMA for waveguides written with increased effective speed and pulse energy. It can be seen that, given a constant pulse energy, an increase in effective writing speed up to $v_{eff} = 2\,\textrm{mm/s}$ ensures a transition back to a single-mode operation regime. As the pulse energy increases, the single-mode operation boundary  shifts toward higher effective speed values.
    
    In most cases, the aspect ratio remained above $0.9$, and it can be brought closer to unity by slightly changing the displacement between writing scans. Simultaneously, the EMA of single-mode multiscan waveguides became larger compared to RWG, since the MFD along the horizontal direction increased and became almost equal to the vertical one. 
    It is worth noting that our goal was to achieve a nearly isotropic coupling between the waveguides. Therefore, it is not the EMA itself that should be taken into account, but rather the coupling constant, which is the integral of the mode field and the refractive index profile \cite{snyder2012optical}. As will be shown below, even for the effective writing speed of $8\,\textrm{mm/s}$ the coupling constant was maintained at a value not exceeding the maximum coupling strength for RWG.

\subsection{\label{sec:anisotropy} Coupling anisotropy}

    \begin{figure*}
        \includegraphics[width=1.0\linewidth]{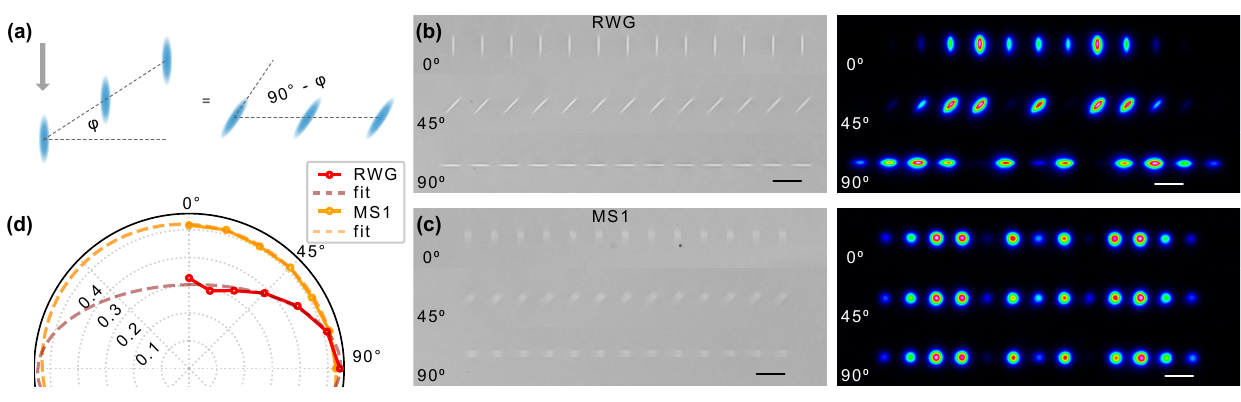}
        \caption{\label{fig_3}
        (a) Schematic representation of waveguides written at different depths, forming a one-dimensional lattice. The angle $\varphi$ denotes the tilt angle of the lattice relative to the horizontal line. Tilting the lattice is equivalent to tilting the waveguides in a horizontal lattice. 
        (b) Left: microscopic view of the 1D lattices consisting of $13$ RWGs and tilted by different angles (the original images are rotated for convenience). The lattice period is $30\,\mu\textrm{m}$. Right: the output DDPs corresponding to central channel excitation. Scale bars correspond to $30\,\mu\textrm{m}$.
        (c) Same as (b), but for MS1 waveguides. The lattice period is chosen slightly lower ($26\,\mu\textrm{m}$) to bring the coupling constants closer to the maximum value obtained for the RWG.
        (d) Measured dependencies of the coupling constant on the lattice tilt angle for RWG and MS1 waveguides. The dotted lines show fits to the data using the polar representation of an ellipse.
        }
    \end{figure*}
    
    In this section, we examine the coupling anisotropy inherent to RWGs and multiscan waveguides MS1 ($N = 8$, $s = 1.3\,\mu\textrm{m}$, $v_{eff} = 6\,\textrm{mm/s}$, $E_p = 270\,\textrm{nJ}$). When two RWGs are written at different depths, their cross-sections turn out to be tilted at an angle $\varphi$ relative to the line connecting the waveguide centers (Fig.~\ref{fig_3}(a)). The coupling constant is given by the overlap integral (see \ref{sec:Methods}), and consequently depends on $\varphi$. Since RWGs are characterized by a strongly elongated cross-section and an elliptical eigenmode, the angular dependence of $C$ becomes pronounced. The multiscan approach enables the realization of waveguides with nearly circular eigenmodes, which in turn reduces coupling anisotropy.
    
    To characterize the angular dependence of the coupling constant, we fabricated one-dimensional waveguide lattices with fixed nearest-neighbor spacing ($30\,\mu\textrm{m}$ for RWGs and $26\,\mu\textrm{m}$ for MS1 waveguides) and different vertical offsets between adjacent waveguides, thereby varying the lattice orientation (Fig.~\ref{fig_3}(b, c)). The absolute value of the coupling constant $C$ was determined from the discrete diffraction pattern (DDP) in one-dimensional waveguide lattice (see \hyperref[sec:Methods]{\textbf{Methods}}). The measured angular dependencies of the coupling constant were fitted with the polar equation of an ellipse (Fig.~\ref{fig_3}(d)), and the coupling anisotropy $\Delta_{C}$ was quantified as the normalized difference between the major and minor semiaxes of the ellipse (see \hyperref[sec:Methods]{\textbf{Methods}}). The coupling anisotropy for MS1 waveguides $\Delta_{C}^{MS1} = 0.05 \pm 0.01$ is significantly reduced compared to that for RWG ($\Delta_{C}^{RWG} = 0.45 \pm 0.02$).

    Notably, despite the significant reduction in coupling anisotropy, the waveguide itself remains anisotropic. This behavior results from both the intrinsic structure of the refractive index profile (Fig.~\ref{fig_1}) and the material stress accumulated during the writing process \cite{corrielli2018symmetric}. Consequently, the lowest coupling anisotropy is observed for a rectangular cross‑section rather than a square one.

\subsection{\label{sec:square_lattice} Application to 3D waveguide arrays}

    \begin{figure*}[t]
        \includegraphics[width=1.0\linewidth]{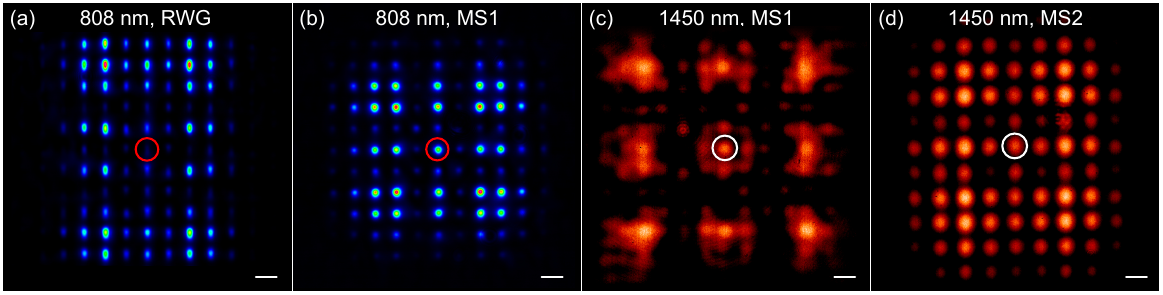}
        \caption{\label{fig_4} 
        Comparison of DDP corresponding to the excitation of the central waveguide in square lattices composed of $11 \times 11$ waveguides. 
        (a) DDP at wavelength of $808\,\textrm{nm}$ in RWG lattice with a nearest-neighbor spacing of $30\,\mu\textrm{m}$ in both directions. 
        (b) Same as panel a), but for MS1 waveguide lattice.
        (c) DDP at a wavelength of $1450\,\textrm{nm}$ in the same lattice as shown on (b).
        (d) DDP at a wavelength of $1450\,\textrm{nm}$ in MS2 waveguide lattice with a nearest-neighbor spacing of $35\,\mu\textrm{m}$.
        Scale bars correspond to $30\,\mu\textrm{m}$.
        }
    \end{figure*}

    In this section, we consider square arrays of equally spaced waveguides to demonstrate the suitability of the developed multiscan writing regime for creating three-dimensional waveguide arrays with nearly isotropic DDP. In such a system, excitation of the central channel is expected to yield a symmetrical DDP. In turn, nonuniformities of the waveguides within the array will cause distortion of DDP. An important source of fluctuations that can limit the use of multiscan for writing three-dimensional waveguide structures is long-term instabilities of optical and positioning systems (see \hyperref[sec:Discussion]{\textbf{Discussion}}). Our multiscan regime enables successful fabrication of such structures due to the increased writing speed.

    Fig.~\ref{fig_4}(a,b) shows the DDP in square waveguide lattices composed of RWGs and MS1 waveguides. In the first case, the DDP exhibits a significant degree of anisotropy, while in the second case, the DDP is identical for horizontal and vertical directions.

    As the laser wavelength increases, the V-number decreases rapidly (Appendix~\ref{appendix:V_number}), resulting in a strong increase in EMA and, consequently, in coupling constant, leading to the need for an increased spacing between the waveguides (Fig.~\ref{fig_4}(c)). Simultaneously, the range of depths over which spherical aberrations have a negligible effect on waveguide performance is restricted. Multiscan is a versatile approach to waveguide design, offering enhanced control over eigenmode. Increasing the number of scans may reduce the EMA at small displacements between adjacent scans, as shown in section~\ref{sec:multiscan}. For a wavelength of $808\,\textrm{nm}$, multiscan waveguides typically exhibit multimode behavior, while at a wavelength of $1450\,\textrm{nm}$ they have a well-localized eigenmode in contrast to the RWG. Fig.~\ref{fig_4}(d) shows a DDP in a square lattice at a wavelength of $1450\,\textrm{nm}$ composed of multiscan waveguides MS2 ($N = 16$, $s = 0.7\,\mu\textrm{m}$, $v_{eff} = 4\,\textrm{mm/s}$, $E_p = 350\,\textrm{nJ}$). Table~\ref{tab:parameters} summarizes the writing parameters for the RWG, MS1, and MS2 waveguides.
    
    \begin{table}[h]
        \caption{\label{tab:parameters} Writing parameters of the RWG, MS1, and MS2 waveguides.
        }
        \begin{ruledtabular}
            \begin{tabular}{c@{\hspace{2mm}} c c@{\hspace{5mm}} c@{\hspace{2mm}} c@{\hspace{2mm}} c c}
                &
                $N$ &
                $s,\,\mu\textrm{m}$ &
                $v_{eff},\,\textrm{mm/s}$ &
                $v_{scan},\,\textrm{mm/s}$ &
                $E_p,\,\textrm{nJ}$ &
                $\lambda,\,\textrm{nm}$ \\
                \colrule
                \textbf{RWG} & 1  &  -   & 1   & 1   & 270 & 808 \\ 
                \textbf{MS1} & 8  & 1.3  & 6   & 48  & 270 & 808 \\ 
                \textbf{MS2} & 16 & 0.7  & 4   & 64   & 350 & 1450 \\ 
            \end{tabular}
        \end{ruledtabular}
    \end{table}

\subsection{\label{sec:losses} Losses}

    We compared the propagation loss (PL) and bending loss (BL) for RWG, MS1 waveguide, and fixed-width waveguides ($12\,\mu\textrm{m}$), see \ref{sec:multiscan}). For more information on loss measurement, see \hyperref[sec:Methods]{\textbf{Methods}}. The RWG has a PL of $0.1 \pm 0.03\,\textrm{dB/cm}$.
    We observed similar values of PL for both MS1 and fixed-width waveguides. Since fixed-width waveguides are few-mode and the EMA of the fundamental mode shrinks as $N$ increases, the BL in multiscan waveguides is progressively reduced with increasing $N$. As for MS1 waveguide, it possesses EMA close to that of RWG. The BL in this case is also similar to the value for RWG.

    \begin{figure}[b]
        \includegraphics[width=1.0\linewidth]{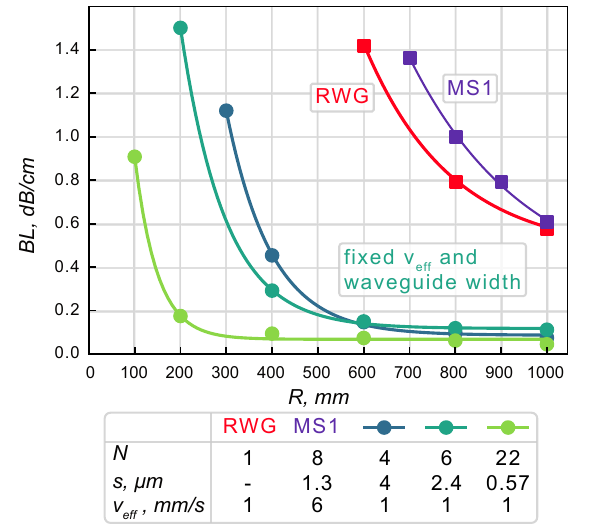}
        \caption{\label{fig_5}
        Measured BL at $808\,\textrm{nm}$ for the RWG (red), the MS1 waveguide (purple), and multiscan waveguides with different $N$, fixed width and writing speed (shades of green). The table under the plot summarizes the parameters of the waveguides shown. Pulse energy was $270\,\textrm{nJ}$ in all cases.
        }    
    \end{figure}

\section{\label{sec:Discussion} Discussion and conclusion}
    
    We found that varying the multiscan parameters while maintaining a fixed effective writing speed enables control not only over the eigenmode aspect ratio but also over the EMA. First, the EMA exhibits a minimum at an optimal displacement between adjacent scans, approximately equal to the width of a single track. Second, for a given waveguide width, the EMA decreases with increasing number of scans under constant-$v_{\mathrm{eff}}$ conditions.

    In the case of a fixed-width waveguide, it is interesting to consider the writing beam distribution averaged over writing time. If the number of scans is large enough for adjacent tracks to overlap, then the time-averaged intensity distribution in the modification region resembles that of a flat-top beam. This observation suggests that single-scan writing with a flat-top beam could provide an efficient way to fabricate similar waveguides at higher speed, since in our approach the speed of each scan increases with the number of scans. Flat-top intensity profiles can be generated using beam-shaping optics or SLM, making this a promising direction for future research. A simpler experimental approach is beam shaping with cylindrical optics, but it retains a Gaussian profile in the focal area. Consequently, for the same pulse energy, the modification size is smaller than that produced by a flat‑top beam. Achieving the same modification size requires a higher peak intensity and a lower writing speed, which may alter the modification type. Thus, multiscan writing provides a simple and efficient way to produce low-aspect-ratio waveguides without increasing the writing time.
    
    It is worth noting that we examined multiscan writing of type-I modifications. First, we were able to accelerate the writing process by up to a factor of eight compared to RWG. Second, we successfully adapted the original writing regime of RWG operating at $808\,\textrm{nm}$ to fabricate waveguides suitable for the telecommunication wavelength range. This was due to the large size of a single modification, which itself possesses the waveguiding properties, as well as the high ellipticity of the single modification cross-section, leading to a rapid increase in the V-number during multiscan. When an individual type-I modification does not exhibit waveguiding properties, the effective writing speed demonstrated in our work cannot be achieved \cite{skryabin2024femtosecond}. The same principle can be applied to type-II modification. In Ref.~\cite{ackermann2025high} high writing speed for waveguides operating at telecom wavelengths was demonstrated using UV writing in fused silica glass. This is likely due to the use of individual modifications with a high aspect ratio, similarly to our work.
    
    We have shown that multiscan can significantly reduce the coupling anisotropy between waveguides, which is beneficial for applications where maintaining the geometric symmetry of the system is crucial. There is a lack of examples in the literature of using multiscan to create 3D waveguide arrays, and those we found are limited to planar waveguide arrays in short samples \cite{ghosh2012ultrafast, skryabin2024femtosecond}. This is likely because such structures are sensitive to the long-term stability of optical and positioning systems during the fabrication process. Although laser writing experiments typically use positioning systems with submicron accuracy, the mechanical stability of the beam preparation system can be a limiting factor, as it may exhibit micron-scale displacements, leading to a corresponding shift of the focal spot within the sample. While in the case of single-scan waveguides such fluctuations have a negligible effect on the propagation constant and inter-waveguide couplings (e.g., the nearest-neighbor spacing was around $30\,\mu\textrm{m}$), in the case of multiscan waveguides even a submicron displacement between adjacent scans significantly affects the mode shape (Fig.~\ref{fig_1}). These errors introduce random modulation of the propagation constant along the propagation direction and become especially pronounced over long fabrication times (i.e., at low writing speeds, for large numbers of waveguides, and for long structures). We successfully applied multiscan writing to fabricate continuously coupled waveguide arrays composed of $\sim 100$ waveguides in a $50\,\textrm{mm}$ long sample. 
    This indicates a high degree of uniformity of multiscan waveguides along the entire sample length. We attribute this to two factors: the high writing speed, which reduced the processing time, and the use of displacements that resulted in only minor overlap between adjacent tracks.

    In conclusion, we have systematically investigated multiscan writing under fixed exposure conditions and shown that the number of scans and their relative displacement provide flexible control over both the size and the aspect ratio of the waveguide eigenmode. The optimized multiscan regime reduces the inter-waveguide coupling anisotropy from $0.45$ to $0.06$, accelerates the fabrication by up to a factor of eight, and extends the original writing regime from $808\,\textrm{nm}$ to telecommunication wavelengths, while maintaining a propagation loss of about $0.1\,\textrm{dB/cm}$. Using this approach, we fabricated uniform $11 \times 11$ waveguide arrays exhibiting nearly isotropic discrete diffraction at both $808\,\textrm{nm}$ and $1450\,\textrm{nm}$. We believe that these results offer a practical route toward large-scale three-dimensional waveguide arrays for topological photonics, where isotropic coupling is essential for preserving lattice symmetry.

\begin{acknowledgments}
    The work was supported by Russian Science Foundation 
    (\href{https://rscf.ru/en/project/22-12-00353/}{grant 22-12-00353-$\Pi$}).
    S.A.Z. acknowledges support by the Foundation for the Advancement of Theoretical Physics and Mathematics “BASIS” (22-2-2-26-1).
\end{acknowledgments}

\bibliography{bibliography}



\appendix

\makeatletter
\renewcommand{\fnum@figure}{\thefigure}
\makeatother

\renewcommand{\thefigure}{S\arabic{figure}}
\setcounter{figure}{0}

\begin{figure*}
    \includegraphics[width=1.0\linewidth]{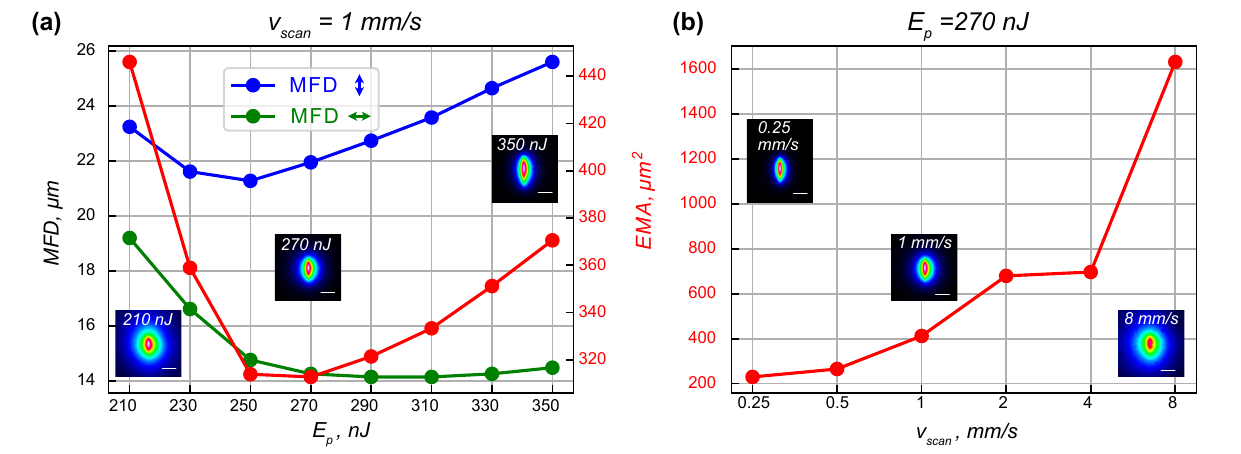}
    \caption{\label{fig:S1}
    (a) Dependence of MFDs and EMA on pulse energy for a single-scan waveguide at writing speed $v = 1\,\textrm{mm/s}$.
    (b) Dependence of EMA on writing speed for pulse energy $E_p = 270\,\textrm{nJ}$.
    }
\end{figure*}

\newpage

\section{\label{appendix:RWG} Reference waveguide}
    
    When selecting the writing parameters for the RWG, we fixed the writing speed $v_{scan} = 1\,\textrm{mm/s}$ and measured the dependencies of MFDs and EMA on pulse energy $E_p$ (Fig.~\ref{fig:S1}(a)). It can be seen that as $E_p$ increase, the MFD in the horizontal direction reaches saturation, while the MFD in the vertical direction reaches a minimum and then starts to grow. This occurs because, at higher $E_p$, the refractive index contrast $\Delta n$ reaches a certain maximum value $\Delta n_{max}$, but at high energies the modification becomes larger in the vertical direction. $E_p = 270\,\textrm{nJ}$ was chosen for RWG, for which the minimum EMA was observed. Fig.~\ref{fig:S1}(b) shows the dependence of EMA on writing speed $v_{scan}$. Since EMA grows rapidly with writing speed, we chose $v_{scan} = 1\,\textrm{mm/s}$ for RWG as a compromise between the resulting writing time and eigenmode localization.

\section{\label{appendix:scan_ordering} Scan ordering}
    
    \begin{figure}[b]
        \includegraphics[width=1.0\linewidth]{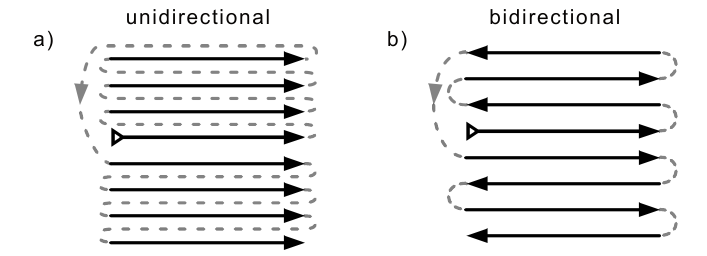}
        \caption{\label{fig:S2}
        Two scanning strategies used in this work. 
        (a) Unidirectional: all scans are written in the same direction. 
        (b) Bidirectional: the writing direction alternates between successive scans. Solid lines indicate writing scans, during which the laser beam is focused inside the sample and material modification occurs; dashed lines indicate the non-writing segments of the writing routine.
        }
    \end{figure}
    
    The order and direction in which scans are written within a multiscan waveguide can significantly affect the resulting mechanical stress distribution and the shape of its eigenmode \cite{sun2020control, lee2021low, skryabin2024femtosecond}. In this work, all scans within a waveguide were written from the center toward the edge in two halves: scans farther from the center were assigned higher indices, and once the first half was completed, writing resumed from the center and proceeded in the opposite direction.
    
    Figure~\ref{fig:S2} shows the two scanning strategies used in this work. In the unidirectional strategy, all scans are written in the same direction, whereas in the bidirectional strategy the writing direction alternates between successive scans, eliminating the non-writing sample translations required to return to the starting position. We used the bidirectional strategy primarily when the spatial overlap between adjacent scans was small ($s > 1\,\mu\textrm{m}$); in this regime, the choice of strategy affects the writing time but has almost no effect on the EMA. Nevertheless, we observed that the unidirectional strategy yields a slightly smaller EMA for otherwise identical writing parameters.

\section{\label{appendix:V_number} V-number and refractive index volume}
   
    The waveguide parameter (normalized frequency, or V-number) is a dimensionless quantity used in the analysis of waveguide properties. For a circular step-index waveguide (CSIW), it is defined as follows \cite{snyder2012optical}:
    \begin{equation} \label{V_equation}
        V = \frac{2 \pi}{\lambda} \rho \sqrt{n_{\textrm{co}}^{2} - n_{\textrm{cl}}^{2}} = k \rho \textrm{NA},
    \end{equation}
    where $\lambda$ is the vacuum wavelength, $\rho$ is the core radius, $k$ is the wavenumber, $n_{\textrm{co}}$ and $n_{\textrm{cl}}$ are the refractive indices of the core and cladding, respectively, and $\textrm{NA}$ is the numerical aperture of the waveguide.
    
    Another important parameter is the refractive index profile volume $\Omega$. If $n^2(x, y) = n^2_{cl} + (n^2_{co}-n^2_{cl})\cdot f(x,y)$, where $f(0) = 1$, $f(\infty) = 0$, then
    \begin{equation} \label{ri_volume}
        \Omega = \int_{A_{\infty}} f(x,y)\ dXdY
    \end{equation}
    where $X = x/\rho_x$, $Y = y/\rho_y$, $\rho_x$ and $\rho_y$ - the characteristic core dimensions.
    
    The V-number determines both the single-mode cutoff $V_c$ as well as the degree of mode confinement characterized by the ratio $r_0/\rho$. For a CSIW, $V_c = 2.405$, and the mode field radius is given by $r_0/\rho \approx 1/\sqrt{\ln(V)}$, where $r_0$ denotes the $1/e^2$ intensity radius \cite{snyder2012optical}.
    
    In the case of a circular graded-index waveguide, the $V$-number is defined analogously to Eq.~(\ref{V_equation}), with $n_{\text{co}}$ corresponding to the maximum refractive index in the core. The intensity distribution of the fundamental mode and the cutoff frequency are weakly sensitive to the shape of the refractive-index profile and are largely determined by its volume. They can be estimated using a CSIW with an equivalent profile volume and the same values of $n_{\text{co}}$ and $n_{\text{cl}}$, for which $V_{\Omega} = V \cdot \sqrt{\frac{\Omega}{\pi}}$. If the waveguide is not axially symmetric, the parameters $V_{x}$ and $V_{y}$ are introduced to characterize the mode confinement along each transverse direction.

\section{\label{appendix:zero_displacement} Multiscan with zero displacement}

    \begin{figure}[b]
        \includegraphics[width=1.0\linewidth]{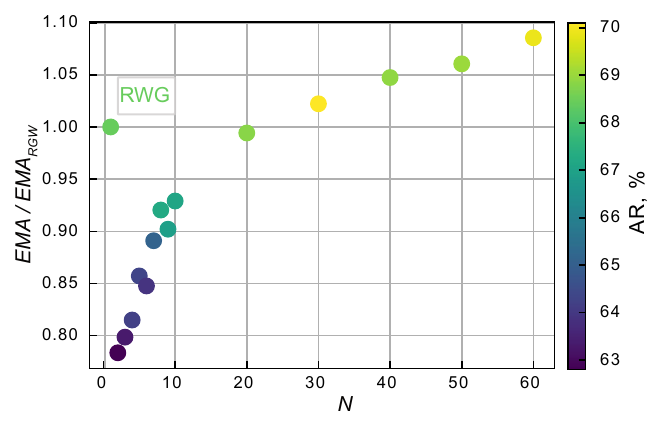}
        \caption{\label{fig:S3}
        Dependence of the normalized EMA on the number of scans at fixed values of $v_{eff}$, $E_{p}$ and $s = 0\,\mu\textrm{m}$. Color shows the aspect ratio of the eigenmode.
        }
    \end{figure}
    
    If the effective writing speed is fixed, and $s = 0\,\mu\textrm{m}$, then a change in $N$ means that a fixed number of pulses is distributed along the length of the waveguide in a different way. The corresponding dependence of EMA on $N$ is shown in Fig.~\ref{fig:S3}. In most cases the EMA was lower than for the RWG, and the minimum EMA was observed at $N = 2$, however, the value of $N$ at which the minimum EMA occurs may depend on the effective writing speed. This approach is conceptually similar to the coordinated change in repetition rate and writing speed used to keep the number of pulses per unit length fixed \cite{ross2024low}. Nevertheless, in our case, the pulse redistribution over time occurs differently, so the writing time of an individual waveguide remains practically unchanged.

\section{\label{appendix:small_s_range} Optimal displacement for multiscan}
    
\begin{figure}[h]
        \includegraphics[width=1.0\linewidth]{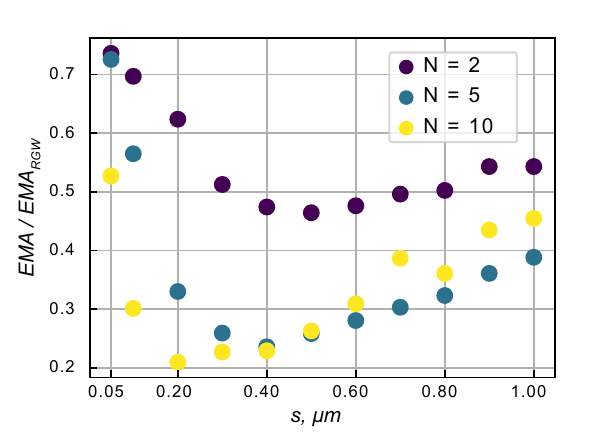}
        \caption{\label{fig:S4}
        Dependence of the normalized EMA on the displacement between scans for different numbers of scans.
        }
    \end{figure}

    In section \ref{sec:multiscan} we investigated the dependence of the EMA on the number of scans and the displacement between them. It was found that in the range of displacements from $0$ to $1\,\mu\textrm{m}$ the EMA reaches its minimum for all $N$. We examined this dependence in more detail in the vicinity of the EMA minimum for several values of $N$. The results are shown in Fig.~\ref{fig:S4}. It can be seen that the displacement corresponding to the EMA minimum has a weak dependence on $N$.

\newpage

\end{document}